\documentclass[journal]{IEEEtran}
\usepackage[figuresright]{rotating}
\usepackage{amssymb}
\usepackage{amsthm}
\usepackage{multicol}
\usepackage{subfigure}
\usepackage{graphicx}
\usepackage{epstopdf}
\usepackage{fullpage}
\usepackage{latexsym,amsmath}
\usepackage{stmaryrd}
\usepackage{algorithm,algorithmic}
\usepackage{subfigure}
\usepackage{amsfonts}
\usepackage{xcolor,multirow}
\usepackage{cite}
\usepackage{bm}
\usepackage{url}
\usepackage{diagbox}
\usepackage{array}
\usepackage{booktabs}
\usepackage{footmisc} 
\usepackage{pifont}

\newtheorem{definition}{Definition}

\ifCLASSINFOpdf
\else
\fi
\begin{document}
	
\title{Constrained low-rank quaternion approximation for color image denoising by bilateral random projections}

\author{Jifei~Miao and Kit~Ian~Kou

\thanks{The authors are with the Department of Mathematics, Faculty of Science
	and Technology, University of Macau, Macau 999078, China  (e-mail: jifmiao@163.com;
	kikou@umac.mo)}}

\markboth{Journal of \LaTeX\ Class Files,~Vol.~14, No.~8, August~2019}%
{Shell \MakeLowercase{\textit{et al.}}: Bare Demo of IEEEtran.cls for IEEE Journals}

\maketitle

\begin{abstract}
In this letter, we propose a novel low-rank quaternion approximation (LRQA) model by  directly constraining the quaternion rank prior for effectively removing the noise in color images. The LRQA model treats the color image holistically rather than independently for the color space components, thus it can fully utilize the high correlation among RGB channels. We design an iterative algorithm by using quaternion bilateral random projections (Q-BRP) to efficiently optimize the proposed model. The main advantage of Q-BRP is that the approximation of the low-rank quaternion matrix can be obtained quite accurately in an inexpensive way. Furthermore, color image denoising is further based on nonlocal self-similarity (NSS) prior. The experimental results on color image denoising illustrate the effectiveness and superiority of the proposed method.

\end{abstract}

\IEEEpeerreviewmaketitle

\section{Introduction}
\IEEEPARstart{I}mage denoising is an important task in the field of image processing. Among numerous image denoising  methods, low-rank matrix approximation  (LRMA) methods have made a great success \cite{DBLP:journals/spl/ParekhS16,DBLP:conf/cvpr/GuZZF14,DBLP:journals/tip/XieGLZZZ16,DBLP:journals/csur/ZhouYZY14}. These methods generally adopt certain rank approximation regularizers, \emph{e.g.}, the nuclear norm which has been proven the tightest convex relaxation of the NP-hard rank minimization function \cite{DBLP:journals/focm/CandesR09}. To better approximate the rank function, the authors in \cite{DBLP:conf/cvpr/GuZZF14} considered assigning different weights to different singular values and proposed the weighted nuclear norm minimization (WNNM) algorithm. Although most of the existing LRMA methods have achieved excellent performance for grayscale image denoising, when handling color images, they may suffer from performance degradation. These methods are inherently designed for grayscale image denoising, when extending them to color images, they usually processes each color channel independently using the monochromatic model or processes the concatenation of three color channels using the concatenation model \cite{DBLP:journals/tip/MairalES08,DBLP:journals/tip/ChenXZ20}. However, these two schemes ignore the inter-relationship among RGB channels, thus they may produce hue distortions in the reconstruction results \cite{DBLP:journals/ijon/YuZY19}.

Recently, quaternion, as an elegant color image representation tool, has achieved great success for color image processing \cite{DBLP:journals/spl/Li13,DBLP:journals/spl/LiWS15,DBLP:journals/spl/PedoneBFH15}. Adopting quaternion algebra, a color image encoded as a pure quaternion matrix is processed holistically and the coupling between the color channels is handled naturally \cite{DBLP:journals/iet-ipr/ChenLSLS14}. More
recently, the authors in \cite{DBLP:journals/ijon/YuZY19} and \cite{DBLP:journals/tip/ChenXZ20} extended the traditional LRMA
methods to quaternion field, and respectively proposed LRQA (including nuclear norm, Laplace function, Geman function, and weighted Schatten norm) and quaternion weighted nuclear norm minimization (QWNNM) algorithms. These methods show promising performance for color image denoising. However, these  methods, especially for the large size color images, suffer from the heavy burden of computing quaternion singular value decompositions (QSVD) which are calculated by their equivalent complex matrices with twice sizes. Therefore, these motivate us to develop a novel model with a fast and accurate algorithm for efficient color image denoising.

The main contributions of this letter are summarized as follows:
\begin{itemize}
	\item Representing a color image as a pure quaternion matrix, we propose a novel constrained low-rank quaternion approximation model for effective color image noise removal.
	\item Instead of computing QSVD, we design a quaternion bilateral random projections algorithm (Q-BRP) to efficiently solve the proposed model, which can accurately and quickly approximate the low-rank quaternion matrix.
	\item The proposed method is further based on the nonlocal self-similarity (NSS) \cite{DBLP:journals/tip/DongSL13} prior and used for color image denoising task. The experimental results demonstrate the competitive performance of the proposed approach compared to the state-of-the-art methods in color image denoising task.
\end{itemize}

The remainder of this letter is organized as follows. Section \ref{sec2} briefly introduces some notations and preliminaries for quaternion algebra. Section \ref{sec3} gives the proposed model and algorithm. Section \ref{sec4} provides some experiments to illustrate the performance of our algorithm,
and compare it with several state-of-the-art methods. Finally, some conclusions are drawn in Section \ref{sec5}.

\section{Notations and preliminaries}
\label{sec2}

\subsection{Notations}
In this letter, $\mathbb{R}$ and $\mathbb{H}$ respectively denote the real space and quaternion space. A scalar, a vector, and a matrix are written as $a$, $\mathbf{a}$, and $\mathbf{A}$, respectively. $\dot{a}$,  $\dot{\mathbf{a}}$, and $\dot{\mathbf{A}}$
respectively represent a quaternion scalar, a quaternion vector, and a quaternion matrix. $(\cdot)^{\ast}$, $(\cdot)^{-1}$, and $(\cdot)^{H}$ denote the
conjugation,  inverse, and  conjugate transpose, respectively. $|\cdot|$ and $\|\cdot\|_{F}$ are respectively the modulus and the Frobenius norm.  ${\rm{tr}}\{\cdot\}$ and ${\rm{rank}}(\cdot)$ denote the trace and rank operators, respectively.

\subsection{Basic knowledge of quaternion algebras}
Quaternion space $\mathbb{H}$ was first introduced by W. Hamilton \cite{articleHamilton84} in 1843. A quaternion $\dot{q}\in\mathbb{H}$ is defined as
\begin{equation}\small
\label{equ2}
\dot{q}=q_{0}+q_{1}i+q_{2}j+q_{3}k,
\end{equation}
where $q_{l}\in\mathbb{R}\: (l=0,1,2,3)$, and $i, j, k$ are
imaginary number units and obey the quaternion rules that
\begin{align}\small
\left\{
\begin{array}{lc}
i^{2}=j^{2}=k^{2}=ijk=-1,\\
ij=-ji=k, jk=-kj=i, ki=-ik=j.
\end{array}
\right.
\end{align}
$\dot{q}$ can be decomposed into a real part $\mathfrak{R}(\dot{q}):=q_{0}$ and an imaginary part $\mathfrak{I}(\dot{q}):=q_{1}i+q_{2}j+q_{3}k$ such that $\dot{q}=\mathfrak{R}(\dot{q})+\mathfrak{I}(\dot{q})$.
If the real part $\mathfrak{R}(\dot{q})=0$, $\dot{q}$ is named a pure quaternion. Given two quaternions $\dot{p}$ and $\dot{q}\in\mathbb{H}$, the sum and multiplication of them are respectively
\begin{equation*}\small
\label{eqn1}
\dot{p}+\dot{q}=(p_{0}+q_{0})+(p_{1}+q_{1})i+(p_{2}+q_{2})j+(p_{3}+q_{3})k
\end{equation*}
and
\begin{align*}\small
\label{eqn2}
\dot{p}\dot{q}=&(p_{0}q_{0}-p_{1}q_{1}-p_{2}q_{2}-p_{3}q_{3})\\
&+(p_{0}q_{1}+p_{1}q_{0}+p_{2}q_{3}-p_{3}q_{2})i\\
&+(p_{0}q_{2}-p_{1}q_{3}+p_{2}q_{0}+p_{3}q_{1})j\\
&+(p_{0}q_{3}+p_{1}q_{2}-p_{2}q_{1}+p_{3}q_{0})k.
\end{align*}
It is noticeable that the multiplication of two quaternions is not
commutative so that in general $\dot{p}\dot{q}\neq\dot{q}\dot{p}$.
The conjugate and the modulus of a quaternion $\dot{q}$ are,
respectively, defined as follows:
\begin{align*}\small
\dot{q}^{\ast}&=q_{0}-q_{1}i-q_{2}j-q_{3}k,\\
|\dot{q}|&=\sqrt{\dot{q}\dot{q}^{\ast}}=\sqrt{q_{0}^{2}+q_{1}^{2}+q_{2}^{2}+q_{3}^{2}}.
\end{align*}

Analogously, a quaternion matrix $\dot{\mathbf{Q}}=(\dot{q}_{mn})\in\mathbb{H}^{M\times N}$ is written
as $\dot{\mathbf{Q}}=\mathbf{Q}_{0}+\mathbf{Q}_{1}i+\mathbf{Q}_{2}j+\mathbf{Q}_{3}k$, where $\mathbf{Q}_{l}\in\mathbb{R}^{M\times N}\: (l=0,1,2,3)$, $\dot{\mathbf{Q}}$ is named a pure quaternion matrix when $\mathfrak{R}(\dot{\mathbf{Q}}):=\mathbf{Q}_{0}=\mathbf{0}$. The Frobenius norm of quaternion matrix is defined as
\begin{equation*}\small
\|\dot{\mathbf{Q}}\|_{F}=\sqrt{\sum_{m=1}^{M}\sum_{n=1}^{N}|\dot{q}_{mn}|^{2}}=\sqrt{{\rm{tr}}\{(\dot{\mathbf{Q}})^{H}\dot{\mathbf{Q}}\}}.
\end{equation*}
More details about quaternion algebra can be found in \cite{Zhang1997Quaternions,Girard2007Quaternions} and their references.

\section{Main results}
\label{sec3}
\subsection{The proposed model}
Color images are  represented as pure quaternion matrix. We aim to recover the clear image $\dot{\mathbf{X}}\in\mathbb{H}^{M\times N}$ from its noisy observation
\begin{equation}\small
\label{eq1}
\dot{\mathbf{Y}}=\dot{\mathbf{X}}+\dot{\mathbf{G}},
\end{equation}
where $\dot{\mathbf{G}}$ is assumed to be Gaussian  noise \cite{DBLP:journals/tip/ChenXZ20,DBLP:journals/ijon/YuZY19}. Supposing that $\dot{\mathbf{X}}$ is low-rank with ${\rm{rank}}(\dot{\mathbf{X}})\leq r$, we formulate the following model
\begin{equation}\small
\label{eq2}
\begin{split}
&\mathop{{\rm{min}}}\limits_{\dot{\mathbf{X}}}\ \|\dot{\mathbf{Y}}-\dot{\mathbf{X}}\|_{F}^{2},\\ 
&\text{s.t.},\ {\rm{rank}}(\dot{\mathbf{X}})\leq r. 	
\end{split}
\end{equation}

\subsection{The proposed algorithm}
The optimal $\dot{\mathbf{X}}$ in (\ref{eq2}) can be obtained by truncated QSVD of $\dot{\mathbf{Y}}$.
The QSVD in \cite{DBLP:journals/tip/ChenXZ20} and \cite{DBLP:journals/ijon/YuZY19} was calculated by their equivalent complex matrices with twice sizes, which generally requires ${\rm{\min}}(\mathcal{O}(MN^{2}),\mathcal{O}(M^{2}N))$ flops, thus it is impractical when $\dot{\mathbf{Y}}$ is of large size. To efficiently solve the problem (\ref{eq2}) 
we develop a Q-BRP algorithm, which replaces the truncated QSVD, and significantly reduces the time cost.
\begin{definition}(Q-BRP)
\label{def1}
For $\dot{\mathbf{Q}}\in\mathbb{H}^{M\times N}$ (\emph{w.l.o.g.}, $M>N$), the Q-BRP of $\dot{\mathbf{Q}}$ can be constructed, \emph{i.e.}, $\dot{\mathbf{P}}_{1}=\dot{\mathbf{Q}}\dot{\mathbf{A}}_{1}$, and $\dot{\mathbf{P}}_{2}=\dot{\mathbf{Q}}^{H}\dot{\mathbf{A}}_{2}$, wherein $\dot{\mathbf{A}}_{1}\in\mathbb{H}^{N\times r}$ and $\dot{\mathbf{A}}_{2}\in\mathbb{H}^{M\times r}$ are random quaternion matrices.

\end{definition}
Then, the Q-BRP based $r$ rank approximation of $\dot{\mathbf{Q}}$ is
\begin{equation}
\label{eq3}
\dot{\mathbf{Q}}=\dot{\mathbf{P}}_{1}(\dot{\mathbf{A}}_{2}^{H}\dot{\mathbf{P}}_{1})^{-1}\dot{\mathbf{P}}_{2}^{H}.
\end{equation}
For random quaternion matrices $\dot{\mathbf{A}}_{1}$ and $\dot{\mathbf{A}}_{2}$, we pick matrices with \emph{i.i.d.} Gaussian entries for real and imaginary parts. These two random quaternion matrices are used to iteratively project $\dot{\mathbf{Q}}$ to $r$-dimensional subspaces. And note that $\dot{\mathbf{A}}_{2}^{H}\dot{\mathbf{P}}_{1}$ is invertible with probability one. Step (\ref{eq3}) is an approximation of the truncated QSVD, which is similar to the case  in the real setting as explained in \cite{Fazel2008Compressed}. Then, following the framework in \cite{Zhou2014GoDec}, the \textbf{c}onstrained \textbf{l}ow-\textbf{r}ank \textbf{q}uaternion \textbf{a}pproximation algorithm by \textbf{b}ilateral \textbf{r}andom \textbf{p}rojections (\textbf{CLQA-BRP}) is summarized in TABLE \ref{tab_algorithm1}
\begin{table}[htbp]
	\caption{\textbf{CLQA-BRP} algorithm.}
	\hrule
	\label{tab_algorithm1}
	\begin{algorithmic}[1]
		\REQUIRE $\dot{\mathbf{Y}}$, $r$, maximum iterations $T$.
		\STATE \textbf{Initialize} $\dot{\mathbf{X}}^{0}=\dot{\mathbf{0}}$, iteration index $t=0$.
	\FOR {$t=1:T$} 
	\STATE $\dot{\mathbf{P}}_{1}=\dot{\mathbf{Y}}\dot{\mathbf{A}}_{1}$, $\dot{\mathbf{A}}_{2}=\dot{\mathbf{P}}_{1}$.
	\STATE $\dot{\mathbf{P}}_{2}=\dot{\mathbf{Y}}^{H}\dot{\mathbf{P}}_{1}$, $\dot{\mathbf{P}}_{1}=\dot{\mathbf{Y}}\dot{\mathbf{P}}_{2}$.
	\IF {${\rm{rank}}(\dot{\mathbf{A}}_{2}^{H}\dot{\mathbf{P}}_{1})<r$}
	\STATE $r:={\rm{rank}}(\dot{\mathbf{A}}_{2}^{H}\dot{\mathbf{P}}_{1})$, regenerate the
	random quaternion matrix $\dot{\mathbf{A}}_{1}$, and go to the first step.
	\ENDIF
	\STATE $\dot{\mathbf{X}}^{t}=\dot{\mathbf{P}}_{1}(\dot{\mathbf{A}}_{2}^{H}\dot{\mathbf{P}}_{1})^{-1}\dot{\mathbf{P}}_{2}^{H}$
	\ENDFOR
		\ENSURE $\dot{\mathbf{X}}^{t}$.
	\end{algorithmic}
	\hrule
\end{table}

\textbf{Remark 1}: (\textbf{Computation complexity}) The computation of $\dot{\mathbf{X}}^{t}$ consists of an inverse of an $r\times r$ ($r\ll {\rm{\min}}(M,N)$) quaternion matrix and three quaternion matrix multiplications. Hence, for $\dot{\mathbf{Y}}\in\mathbb{H}^{M\times N}$ with ${\rm{rank}}(\dot{\mathbf{X}})\leq r$, $\mathcal{O}(MNr)$ flops are required to perform Q-BRP, $\mathcal{O}(r^{2}(2M+r)+MNr)$ flops are required to compute $\dot{\mathbf{X}}^{t}$. We can find that the computational complexity is much less than QSVD-based approximation method.

\textbf{Remark 2}: In the denoising task, we let the maximum iterations $T=1$, which is enough since increasing the number of iterations would not noticeably improve the result. 

\subsection{CLQA-BRP for color image denoising}
Except for low-rank prior, we further consider the NSS property for image denoising tasks. We follow the similar procedure of NSS used in \cite{DBLP:journals/tip/ChenXZ20}. Consequently, given a noisy color image $\dot{\mathbf{Y}} \in\mathbb{H}^{M\times N}$, the whole procedure of our color image denoising method is listed as follows:\\
\textbf{Step 1:} Divide the noisy image $\dot{\mathbf{Y}}$ into overlapped patches with size $w\times w$, and vectorize each patch as a quaternion column vector $\dot{\mathbf{y}}_{i}\in\mathbb{H}^{w^2}$, then find its $n$ nearest neighbor patches (including $\dot{\mathbf{y}}_{i}$ itself) within its local searching window.  At last, the $n$ similar patches are stacked as quaternion column vectors of quaternion matrix $\dot{\mathbf{Y}}_{i}\in\mathbb{H}^{w^2\times n}$.\\
\textbf{Step 2:} For each $\dot{\mathbf{Y}}_{i}$, adopt the proposed CLQA-BRP algorithm to estimate the clear color image
patche $\dot{\mathbf{X}}_{i}$.\\
\textbf{Step 3:} Aggregate $\{\dot{\mathbf{X}}_{i}\}$ together to form the final clear image $\dot{\mathbf{X}}$.

Generally, to obtain a better result, several rounds (denoted by $K$) for Step $1$ and Step $2$ are needed before going to Step $3$.

%

\section{Experimental results}
\label{sec4}
\begin{figure}[htbp]
	\centering
	\includegraphics[width=7.5cm,height=1cm]{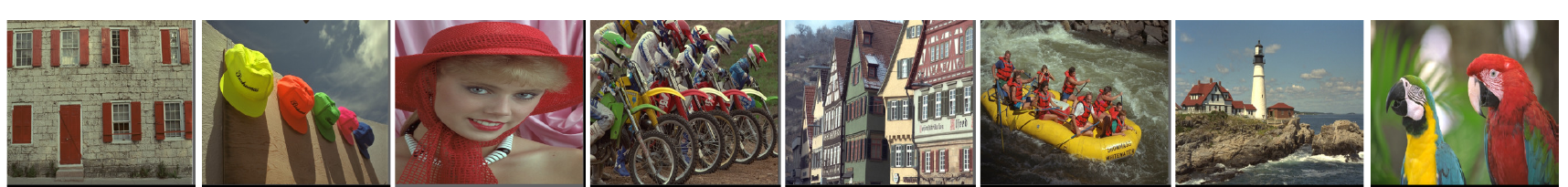}
	\caption{The $8$ color images (from left to right, Image(1) $\sim$ Image
		(8)). are selected form Kodak PhotoCD Dataset (Kodak)\protect\footnotemark[2] all with size $512\times 768\times 3$.}
	\label{Ytu}
\end{figure}

Some experiments on $8$  widely  used  color  images (\emph{see} Fig.\ref{Ytu}) are conducted to evaluate the effectiveness of the proposed method. The additive  white  Gaussian  noise  with  zero  mean  and  variance $\sigma_{n}^{2}$ ($\sigma_{n}=50, 70$ are considered in our experiments) is added to these clean color images.  We compare the proposed method with several latest state-of-the-art methods  including WNNM \cite{DBLP:conf/cvpr/GuZZF14} (a weighted nuclear norm minimization algorithm), QNNM \cite{DBLP:journals/tip/ChenXZ20} (a quaternion nuclear norm minimization algorithm labeled as LRQA-1 in \cite{DBLP:journals/tip/ChenXZ20}), QWNNM \cite{DBLP:journals/ijon/YuZY19} (a quaternion weighted nuclear norm minimization algorithm) and LRQA-WSNN \cite{DBLP:journals/tip/ChenXZ20} (a quaternion weighted Schatten norm minimization algorithm labeled as LRQA-4 in \cite{DBLP:journals/tip/ChenXZ20}). We set the same parameters in the NSS procedure for all the methods. For instance, when $\sigma_{n}=50, 70$, we set patch size to $8\times8$ and $9\times9$ respectively.
The number of similar patches group is set to $120$ and $140$, respectively. The parameters of each
compared algorithm are optimally set or selected as suggested in the source papers.
For the proposed CLQA-BRP algorithm, we set $T=1$, and select $r$ from $\{7,9,15\}$. We employ two widely used quantitative quality indexes (including the peak signal-to-noise ratio (PSNR) and the structure similarity (SSIM)) for performance evaluation. For WNNM, we perform it on each channel of the test color images individually. For quaternion-based methods, we use the same quaternion toolbox\footnote{\url{https://sourceforge.net/projects/qtfm/}}.

\footnotetext[2]{\url{ http://r0k.us/graphics/kodak/}}

\begin{table}[htbp]
	\caption{Quantitative assessment indexes (PSNR/SSIM) of different methods on the eight color images (\textbf{Bold} fonts denote the best performance; \underline{underline} ones represent the second-best results).}
	\centering
	\resizebox{7.8cm}{4cm}{
		\begin{tabular}{|c|c|c|c|c|c|}		
			\hline
			Methods:& WNNM \cite{DBLP:conf/cvpr/GuZZF14} & QNNM \cite{DBLP:journals/tip/ChenXZ20} & QWNNM \cite{DBLP:journals/ijon/YuZY19} & LRQA-WSNN \cite{DBLP:journals/tip/ChenXZ20}  &\textbf{CLQA-BRP} \\ \toprule
			\hline
			Images:  &\multicolumn{5}{c|}{$\sigma_{n}=50$}\\
			\hline
			Image(1)&23.936/0.713&23.958/0.753&24.213/0.751&\underline{24.540}/\underline{0.768}&\textbf{24.633}/\textbf{0.774}\\
			Image(2)&30.025/0.932&29.509/0.923&\underline{30.279}/\underline{0.943}&30.241/\underline{0.943}&\textbf{30.294}/\textbf{0.950}\\
			Image(3)&28.889/0.945&28.321/0.944&\underline{29.033}/\underline{0.953}&29.031/\underline{0.953}&\textbf{29.082}/\textbf{0.959}\\
			Image(4)&24.150/0.780&23.372/0.754&\textbf{24.437}/\textbf{0.793}&24.418/\underline{0.789}&\underline{24.421}/\underline{0.789}\\  
			Image(5)&24.210/0.800&23.500/0.785&24.048/0.798&\underline{24.394}/\underline{0.804}&\textbf{24.436}/\textbf{0.810}\\
			Image(6)&25.368/0.758&24.840/0.756&25.557/\textbf{0.769}&\underline{25.574}/\underline{0.768}&\textbf{25.579}/\underline{0.768}\\
			Image(7)&26.001/0.822&25.392/0.818&\textbf{26.248}/\textbf{0.843}&\underline{26.139}/\underline{0.833}&26.034/0.831\\
			Image(8)&30.130/0.946&29.521/0.939&\textbf{30.356}/\textbf{0.956}&30.211/\underline{0.953}&\underline{30.225}/0.950\\
			\hline
			Aver. &26.589&26.052&26.771&\underline{26.818}&\textbf{26.838}\\ \toprule
			\hline
			Images  &\multicolumn{5}{c|}{$\sigma_{n}=70$}\\
			\hline
			Image(1)&23.137/0.703&22.200/0.672&23.262/0.714&\underline{23.362}/\underline{0.715}&\textbf{23.403}/\textbf{0.720}\\
			Image(2)&28.215/0.913&27.908/0.916&\underline{28.967}/\underline{0.933}&28.586/0.930&\textbf{29.063}/\textbf{0.935}\\
			Image(3)&27.628/0.927&26.563/0.931&\underline{27.959}/\underline{0.944}&27.742/0.938&\textbf{27.982}/\textbf{0.946}\\
			Image(4)&22.490/0.711&21.323/0.654&22.747/\underline{0.717}&\underline{22.863}/\textbf{0.723}&\textbf{22.883}/\textbf{0.723}\\  
			Image(5)&22.557/0.750&21.162/0.685&\textbf{23.070}/\textbf{0.765}&\underline{22.908}/\underline{0.762}&22.875/0.752\\
			Image(6)&24.153/0.714&23.037/0.702&24.294/\underline{0.727}&\underline{24.346}/\textbf{0.729}&\textbf{24.377}/0.724\\
			Image(7)&24.217/0.732&23.734/0.769&\underline{24.702}/0.800&24.701/\underline{0.803}&\textbf{24.717}/\textbf{0.805}\\
			Image(8)&28.624/0.926&27.606/0.928&\textbf{29.136}/\textbf{0.946}&\underline{29.093}/\underline{0.940}&28.992/0.940 \\
			\hline
			Aver. &25.128&24.192&\underline{25.517}&25.450&\textbf{25.537}\\	
			\hline	
	\end{tabular}}
	\label{Index4SR_2}
\end{table}

\begin{figure}[htb]
	\centering
	\subfigure[\scriptsize Original]{
		\includegraphics[width=2.1cm,height=1.5cm]{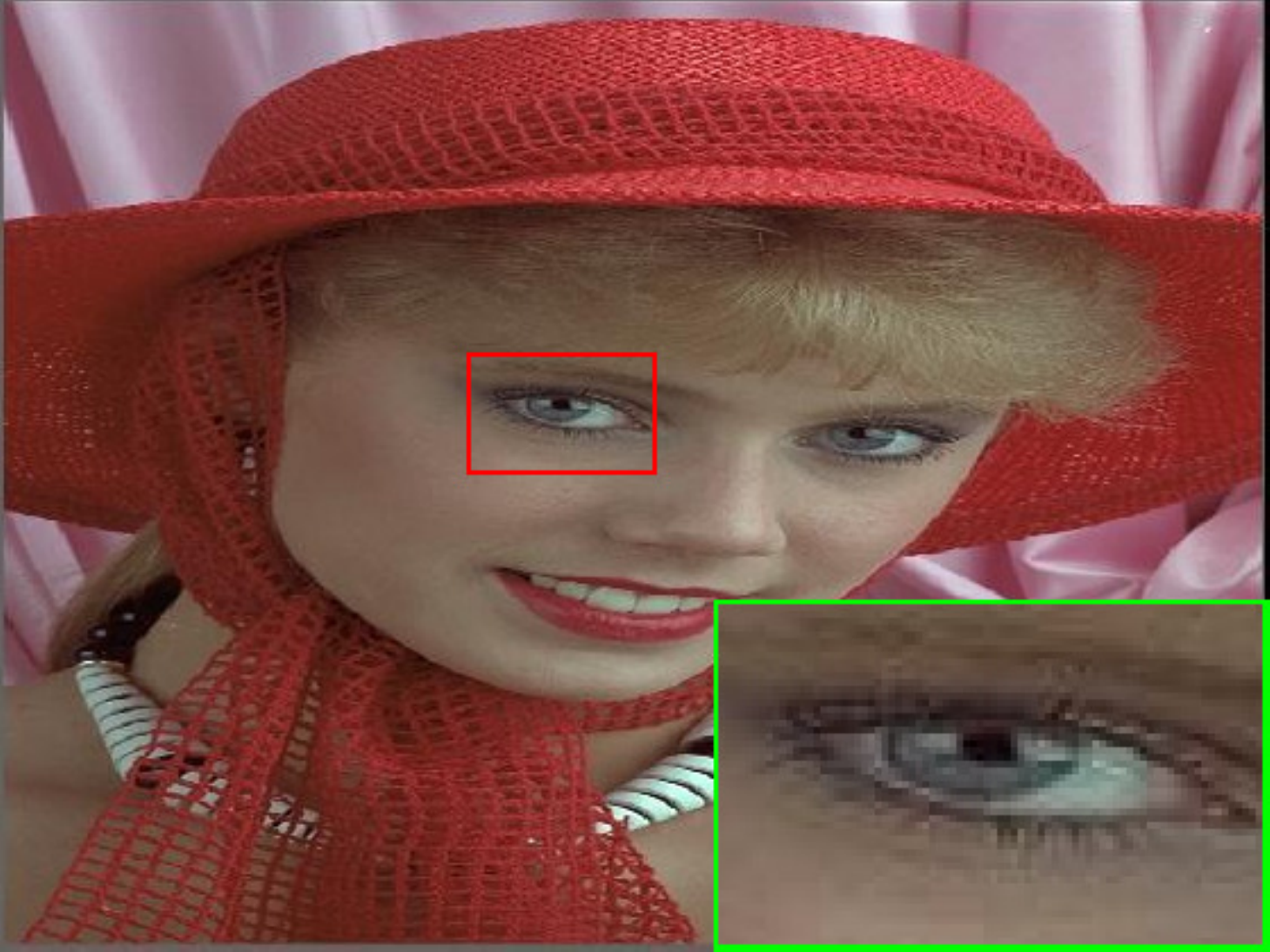}
	}\\
	\hspace{-0.14in}
	\subfigure[\scriptsize Observed]{
		\includegraphics[width=2.1cm,height=1.5cm]{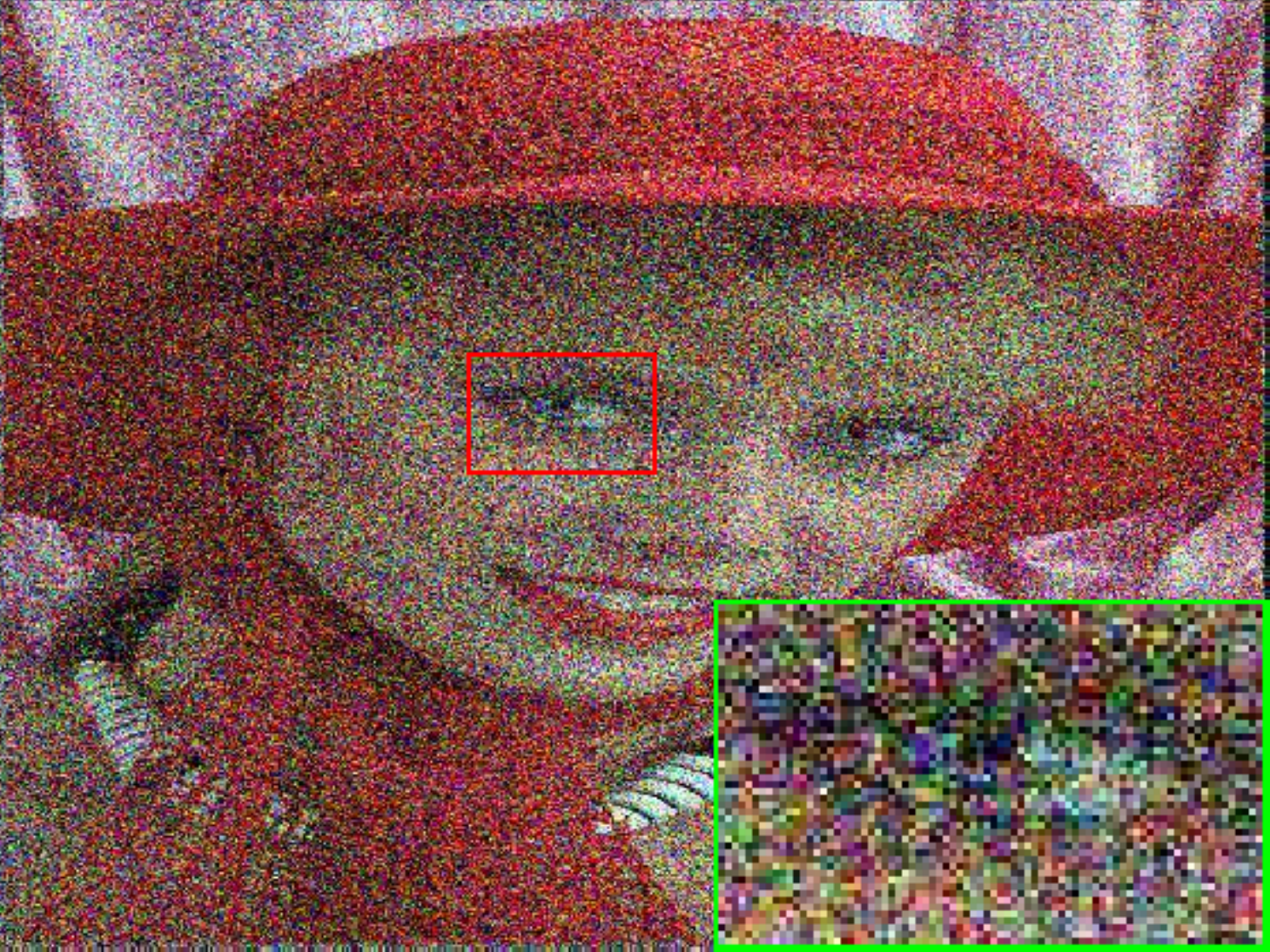}
	}
	\hspace{-0.14in}
	\subfigure[\scriptsize WNNM \cite{DBLP:conf/cvpr/GuZZF14}]{
		\includegraphics[width=2.1cm,height=1.5cm]{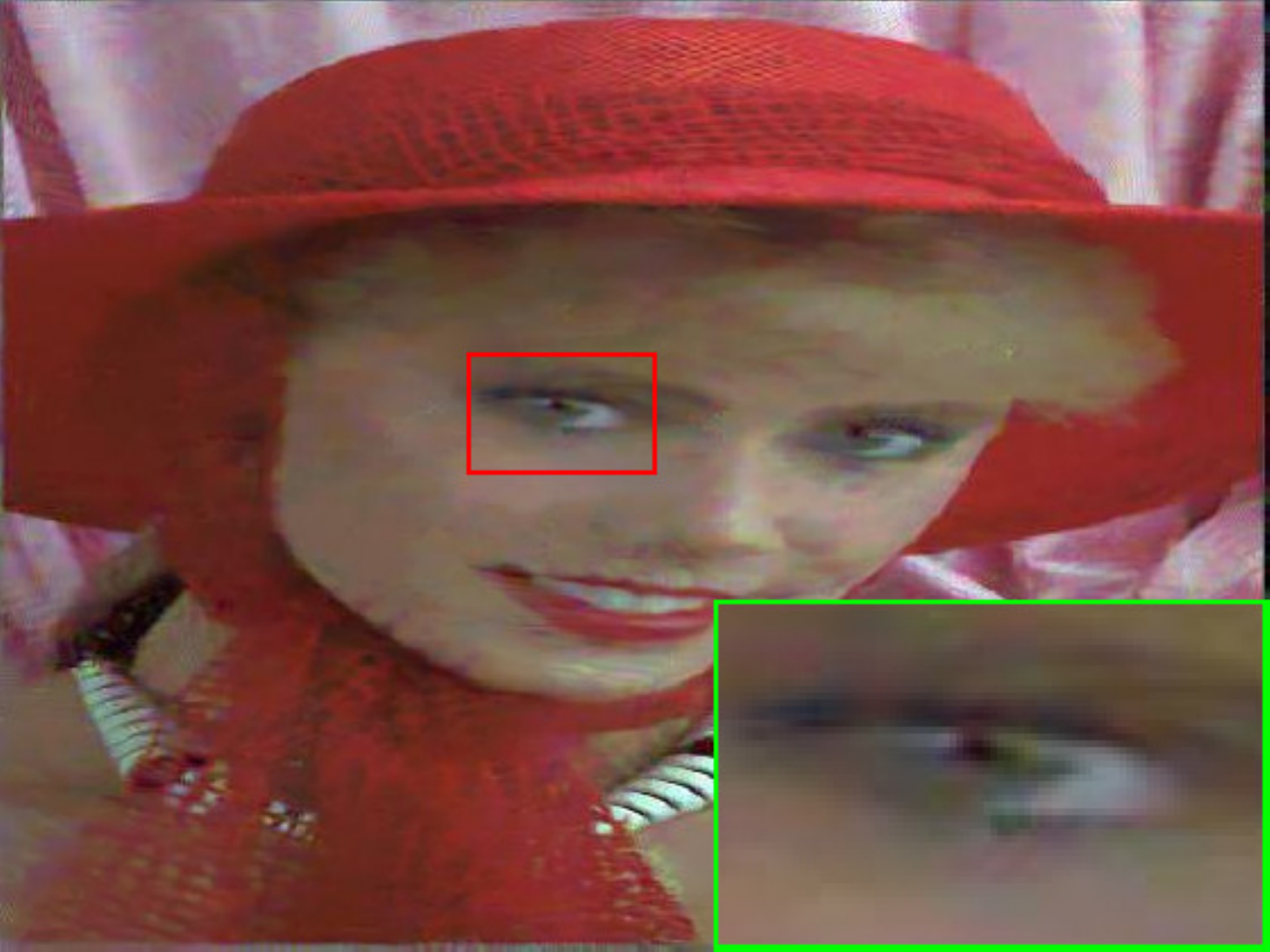}
	}
	\hspace{-0.14in}
	\subfigure[\scriptsize QNNM \cite{DBLP:journals/tip/ChenXZ20}]{
		\includegraphics[width=2.1cm,height=1.5cm]{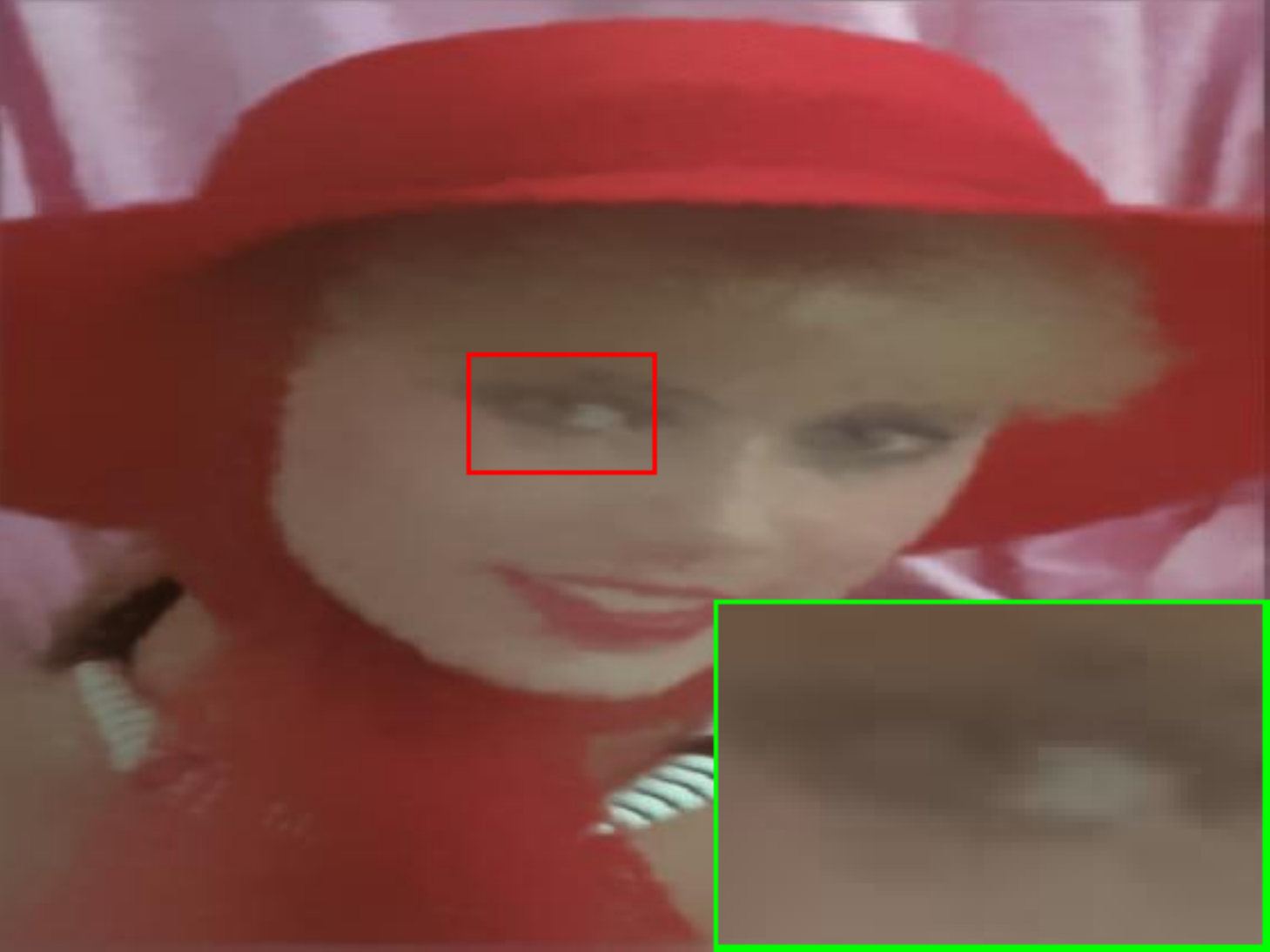}
	}\\
	\hspace{-0.14in}
	\subfigure[\scriptsize QWNNM \cite{DBLP:journals/ijon/YuZY19}]{
		\includegraphics[width=2.1cm,height=1.5cm]{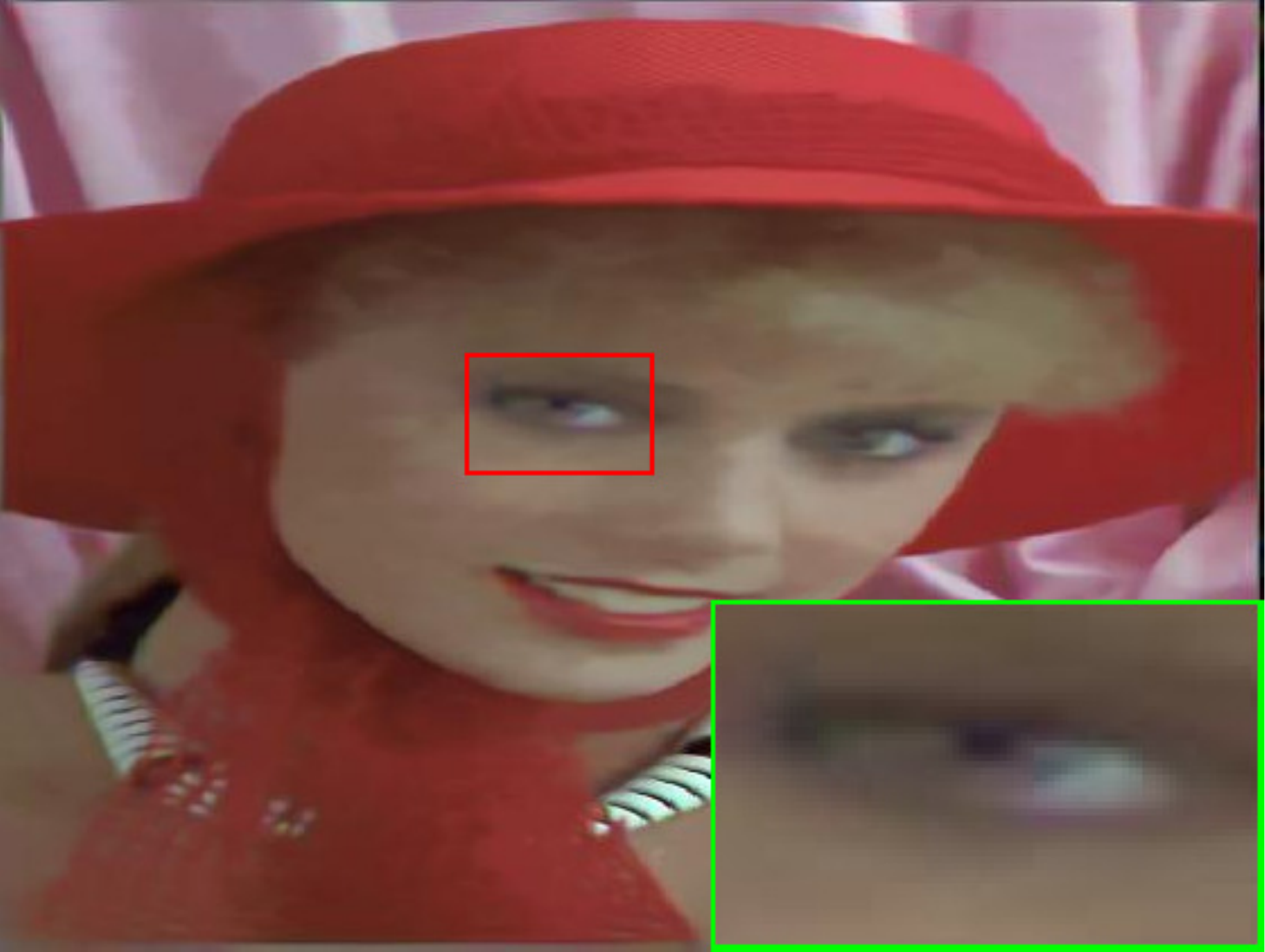}
	}
	\hspace{-0.14in}
	\subfigure[\scriptsize LRQA-WSNN\cite{DBLP:journals/tip/ChenXZ20}]{
		\includegraphics[width=2.1cm,height=1.5cm]{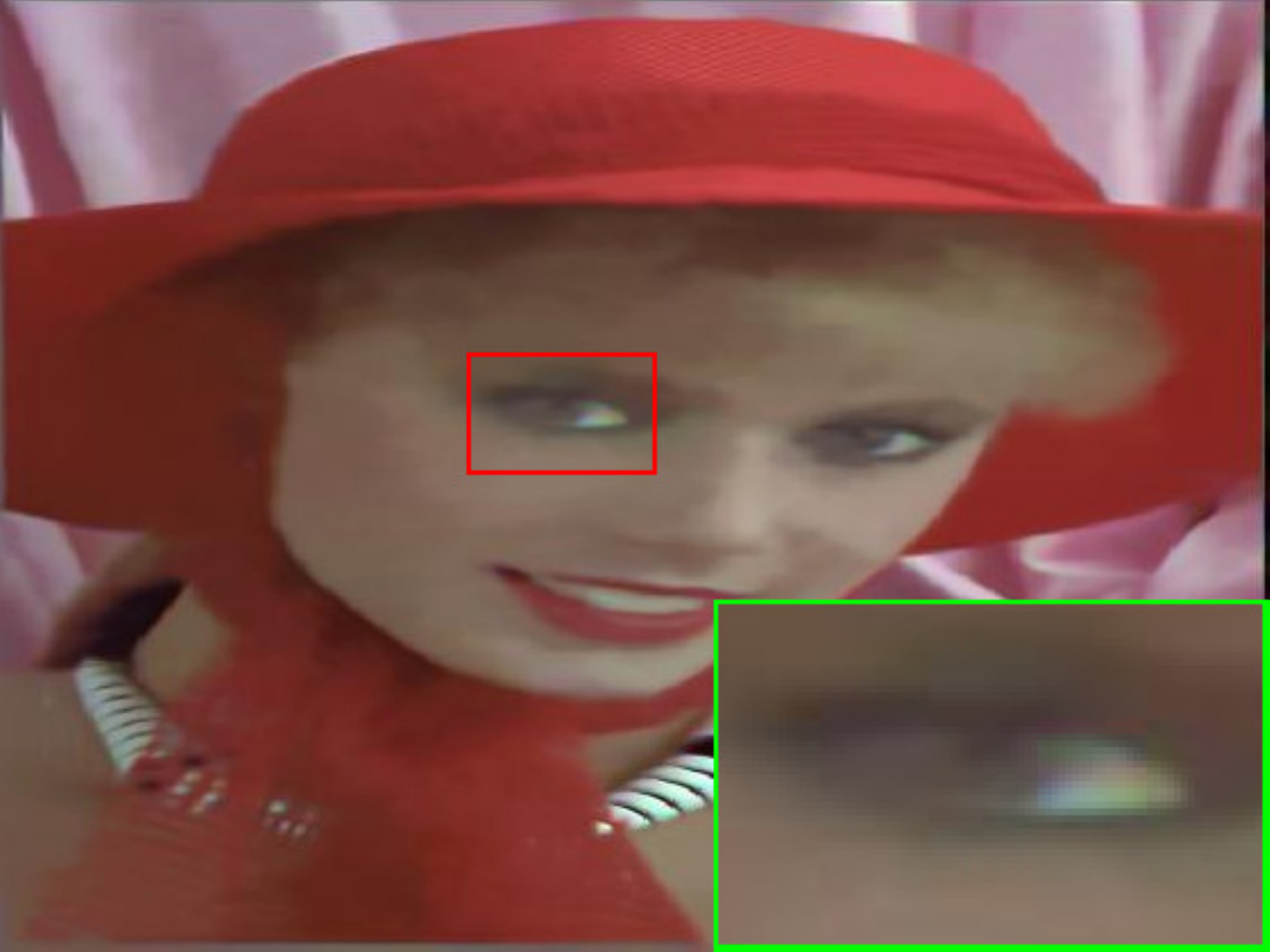}
	}
	\hspace{-0.14in}
	\subfigure[\scriptsize CLQA-BRP]{
		\includegraphics[width=2.1cm,height=1.5cm]{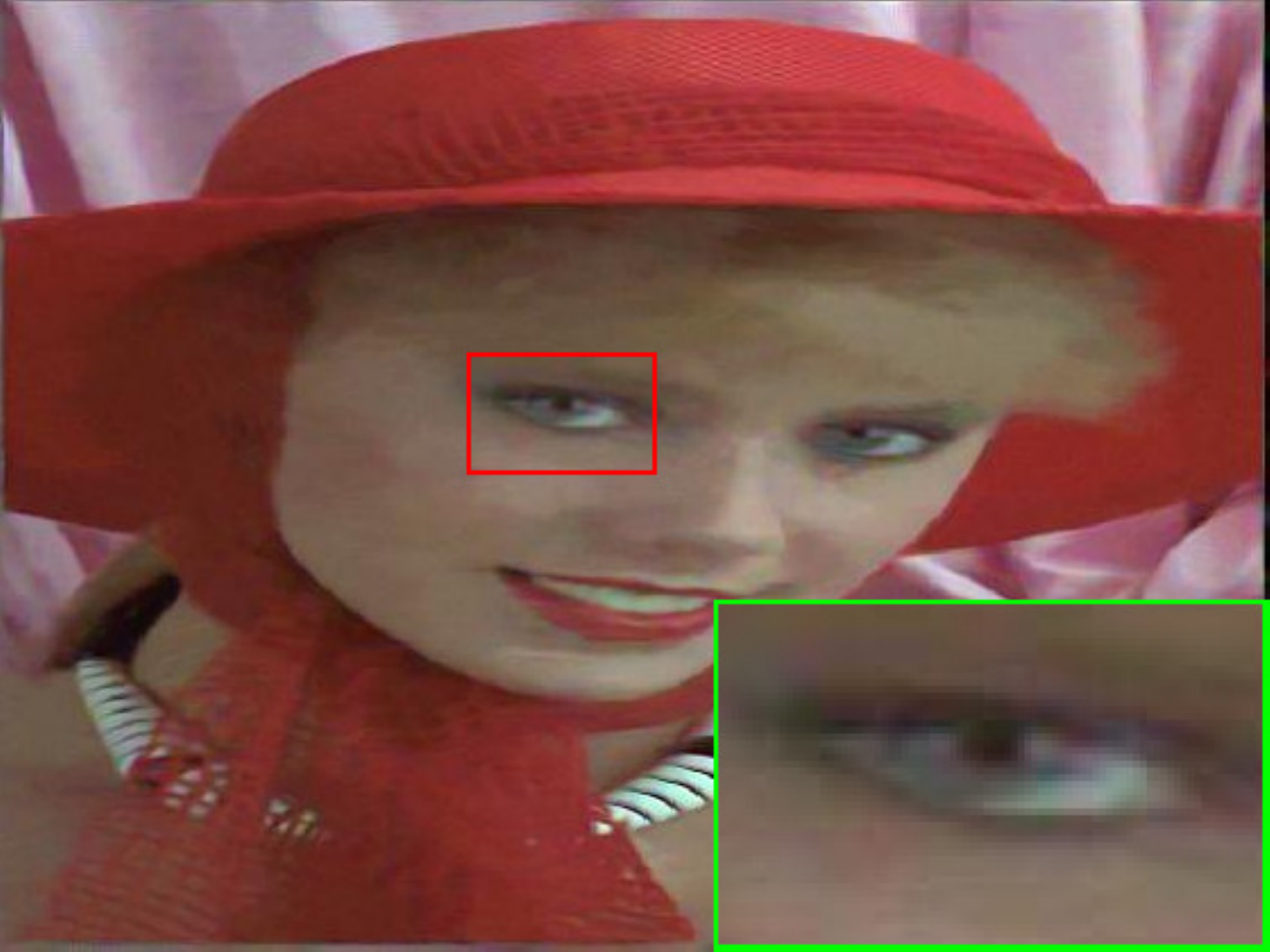}
	}
	\caption{Color image denoising results on Image(3). (a) is the original image. (b) is the observed image ($\sigma_{n}=70$). (c)-(g) are the recovery results of WNNM, QNNM, QWNNM, LRQA-WSNN and CLQA-BRP, respectively. }
	\label{fig01}
\end{figure}
\begin{figure}[!htb]
	\centering
	\subfigure[]{
		\includegraphics[width=8cm,height=3.5cm]{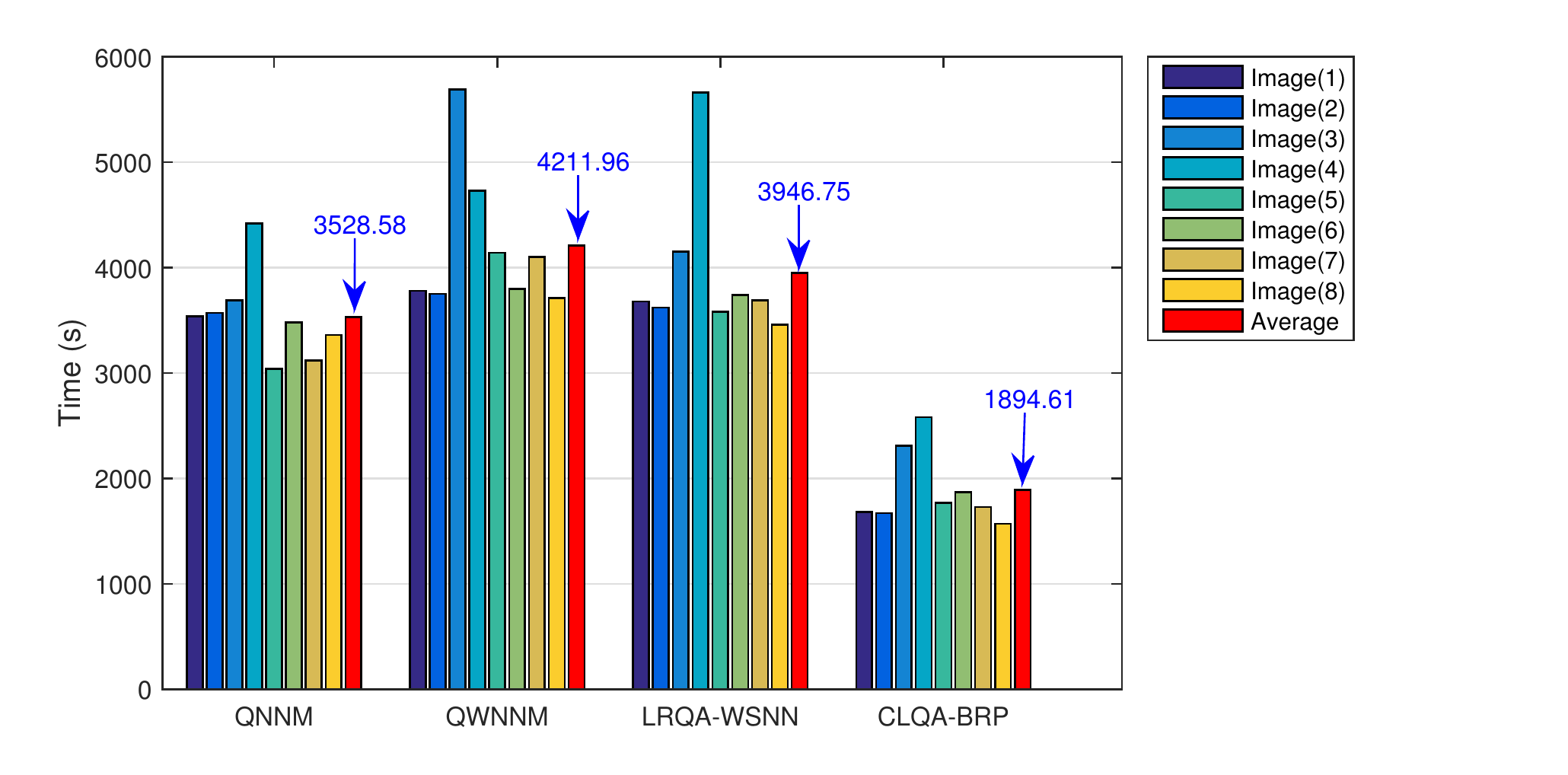}
	}
	\subfigure[]{
		\includegraphics[width=8cm,height=3.5cm]{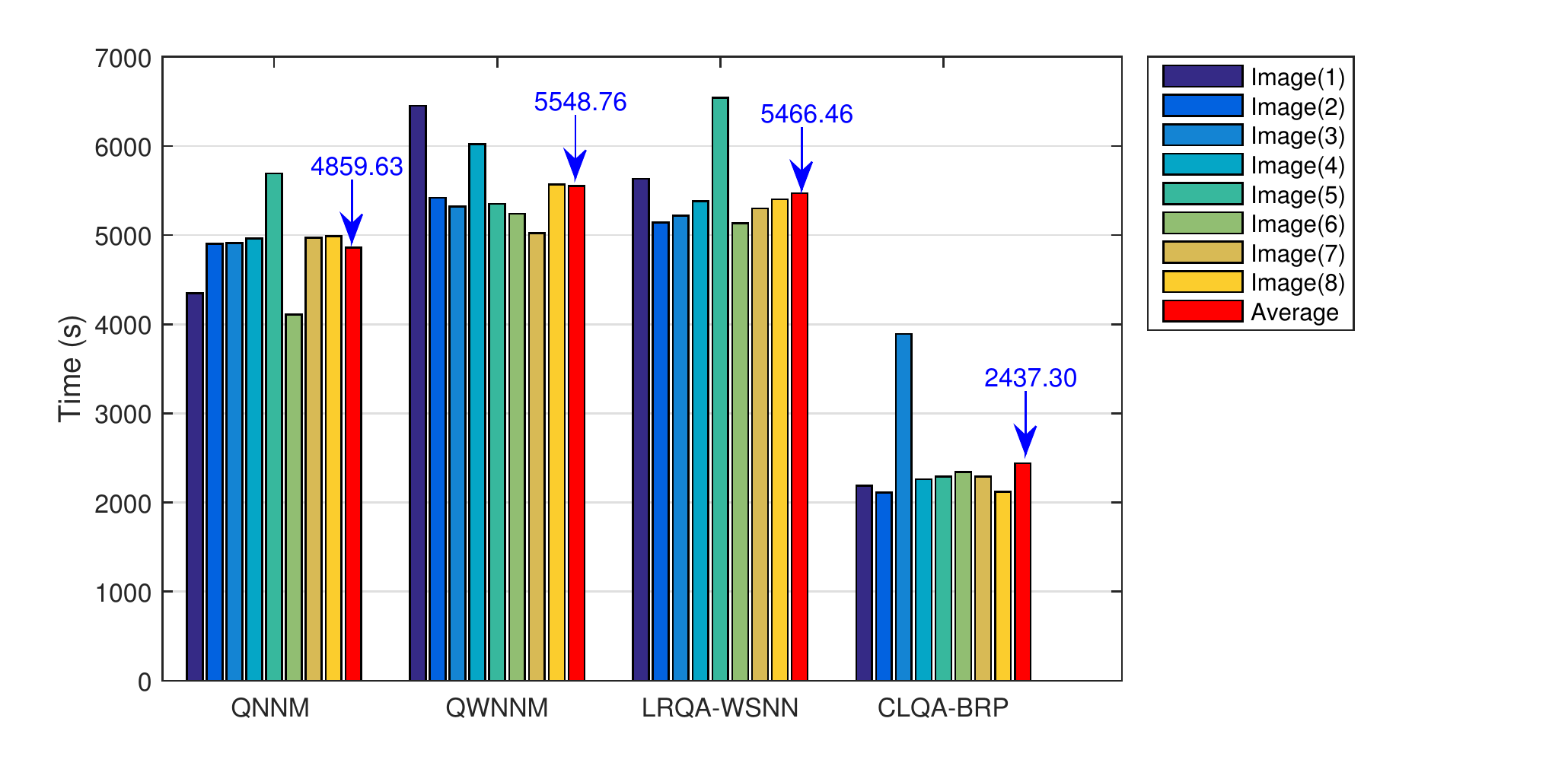}
	}
	\caption{The runtime comparison of all the quaternion-based methods. (a) $\sigma_{n}=50$, (b) $\sigma_{n}=70$.}
	\label{atime}
\end{figure}

TABLE \ref{Index4SR_2} lists the quantitative PSNR and SSIM values (and the average values of PSNR) of all denoising methods. Fig.\ref{fig01} displays the visual comparison between the proposed method and all compared  methods on the Image(3) with $\sigma_{n}=70$. Fig.\ref{atime} shows the runtime comparison of all the quaternion-based methods. According to the obtained results, the following conclusions can be found:
\begin{itemize}
	\item The quaternion-based methods (QWNNM, LRQA-WSNN, and CLQA-BRP) outperform WNNM in all color images. The QNNM has a relatively poor performance, since it, relative to  WNNM, QWNNM, and LRQA-WSNN, does not assign different singular values with different weights.
	\item  The performance of QWNNM, LRQA-WSNN, and our proposed CLQA-BRP are very close, even so, CLQA-BRP has the best average PSNR values in both  $\sigma_{n}=50$ and  $\sigma_{n}=70$.
	\item From Figure \ref{atime}, we can see that the runtime of CLQA-BRP is much shorter than that of other quaternion-based methods  (about half of them). This means that CLQA-BRP is more practical and efficient.
\end{itemize}

\section{Conclusion}
\label{sec5}
We introduced a novel constrained low-rank quaternion approximation model for removing the noise in color images.  Then we design an iterative algorithm by using Q-BRP for efficiently solving the proposed model, which can ignificantly accelerate the approximattion of the low-rank quaternion matrix. Experimental results on color image denoising demonstrate the effectiveness of the proposed CLQA-BRP. In the future, we tend to apply the CLQA-BRP algorithm to color image inpainting and other color image processing tasks.

\section*{Acknowledgment}
This work was supported by The Science and Technology Development Fund, Macau SAR (File no. FDCT/085/2018/A2) and University of Macau (File no. MYRG2019-00039-FST).

\ifCLASSOPTIONcaptionsoff
  \newpage
\fi



%

\bibliographystyle{IEEEtran}
\bibliography{Myreference}

%

%
%
%
%
%
%
%
%
%
\end{document}